\documentclass[article,twocolumn,floatfix]{revtex4}

\usepackage{graphicx} 
\usepackage{subfig}
\usepackage{color}

\newcommand{\ket}[1]{\left| #1 \right\rangle}
\newcommand{\bra}[1]{\left\langle #1 \right|}

\newcommand{\be}{\begin{equation}}
\newcommand{\ee}{\end{equation}}
\newcommand{\bea}{\begin{eqnarray}}
\newcommand{\eea}{\end{eqnarray}}

\usepackage{dcolumn}
\usepackage{bm}
\usepackage{amssymb,amsmath}
\usepackage{color}
\usepackage{float}
\usepackage{tikz}
\usetikzlibrary{arrows}

\definecolor{DarkGreen}{rgb}{0,0.6,0.2}

\begin{document}
\title{How Nonlinear Optical Effects degrade Hong-Ou-Mandel Like Interference}
\author{Imran M. Mirza, S. J. van Enk }
\affiliation{Oregon Center for Optics and Department of Physics\\University of Oregon\\
Eugene, OR 97401}

\begin{abstract}
Two-photon interference effects, such as the Hong-Ou-Mandel (HOM) effect, can be used to characterize to what extent two photons are identical \cite{legero2006characterization}. Furthermore, these interference effects underly linear optics quantum computation. We show here how nonlinear optical effects, such as those mediated by atoms or quantum dots in a cavity, degrade the interference. This implies that, on the one hand, nonlinearities are to be avoided if one wishes to utilize the interference, but on the other hand, one may be able to measure or detect nonlinearities by observing the disappearance of the interference.
\end{abstract}
\maketitle

\section{Introduction}
The Hong-Ou-Mandel (HOM) effect \cite{HOM} is a celebrated example of a pure quantum interference effect. When two photons impinge on two different input ports of a 50/50 beamsplitter, the photons always emerge together in one of the two output ports. The destructive interference between the two paths that lead to the same final state with both photons exiting different output ports can be perfect only if at the output the two photons are indistinguishable. They must, in particular, have identical spectral and polarization states at the output.
In principle there is no such requirement for the photons at the input, and HOM-like interference can occur, for example, between photons of different colors as well \cite{RaymerHOM}, provided there is a frequency-changing mechanism between input and output.

It is straightforward to describe the HOM interference effect in terms of creation operators, one for each electromagnetic field mode. If we denote the two relevant input operators of the 50/50 beamsplitter as $a^\dagger_{{\rm in}}$ and
$b^\dagger_{{\rm in}}$ and the two corresponding output operators by $a^\dagger_{{\rm out}}$ and
$b^\dagger_{{\rm out}}$,  then we may write the effect of the 50/50 beamsplitter as a particular unitary transformation between the pairs of operators:
\bea
\left(
\begin{array}{c}
 a^\dagger_{{\rm in}}       \\
  b^\dagger_{{\rm in}}     
\end{array}
\right)=
\frac{1}{\sqrt{2}}\left(
\begin{array}{cc}
1  & i     \\
i & 1     
\end{array}
\right)\left(
\begin{array}{c}
 a^\dagger_{{\rm out}}       \\
  b^\dagger_{{\rm out}}     
\end{array}
\right).
\eea
This description shows that an input state with two photons in input modes $a_{{\rm in}}$ and $b_{{\rm in}}$ is transformed into an output state of the form
$\frac{((a^\dagger_{{\rm out}})^2-(b^\dagger_{{\rm out}})^2) \ket{{\rm vac}}}{2}$, with
the pair of photons always in a single output mode.
 
Any linear optics setup through which two photons travel effects a
unitary transformation on the mode operators.
The question we consider here is how {\em nonlinear} optics effects affect HOM interference. We consider this question
in the context of coupled cavity arrays. 
Most research on coupled cavity arrays has focused on how classical light can be stored or delayed (there are more than a thousand papers in this area, see for example \cite{yariv1999coupled,heebner2002scissor,heebner2002slow,scheuer2005coupled,baba2008slow,krauss2008we}), but 
such systems will be very useful for  quantum communication purposes, too.
In particular, such cavity arrays can be easily integrated with fiber optics, and they can be used to accurately introduce small time delays of single-photon wavepackets. 
One may expect cavity arrays to be 
used for
entanglement purification protocols and quantum repeaters, which promise to increase the distance over which quantum key distribution can be securely employed \cite{qrepeat,qkd}. This provides some additional motivation for studying this particular physical system.
 
We will include the generation of the two photons explicitly, by assuming we have two single emitters (which could be single atoms or single quantum dots or NV centers in diamond \cite{santori2002indistinguishable,mckeever2004deterministic,englund2010,riedrich2011}), one in each of two cavities (see FIG.~1). 
This kills two birds with one stone:
the two emitters will provide nonlinear optical effects, and the two photons whose interference effects we wish to study are automatically described realistically as wave packets. 
Note that the measurement of non-linearity in the present cavity-QED setup can be performed by following the procedures described in \cite{fink2008climbing, hood2000atom}, which gives our work more experimental feasibility.
The only work we are aware of in more or less the same direction as ours is a paper
\cite{nazir2009overcoming} on the HOM effect in a solid-state setup, with ambient noise taken into account, and with the two emitters included in the description, too (but no cavities, and hence no strong nonlinearities).

We describe our system and the theoretical methods we employ in Section II. The description of unidirectional coupling of two cavities can be done elegantly within the formalism of
quantum cascaded systems combined with quantum trajectories \cite{carmichael1993quantum,gardiner1993driving}. In our case we can still straightforwardly use the latter, but the former theory has to be adjusted to account for bidirectional coupling (so that the photons can travel back and forth between the two cavities). 
With the help of these methods, we study two-photon interference effects in Section III. We simulate there an experiment in which one records which detector(s) detect the two photons, and at what times. The important information is then found in correlations between the two photon detections. 

\section{Two spatially separated atom-cavity systems}
\subsection{Model and Hamiltonian}
\begin{figure*}[t]
\includegraphics[width=5.3in,height=2.16in]{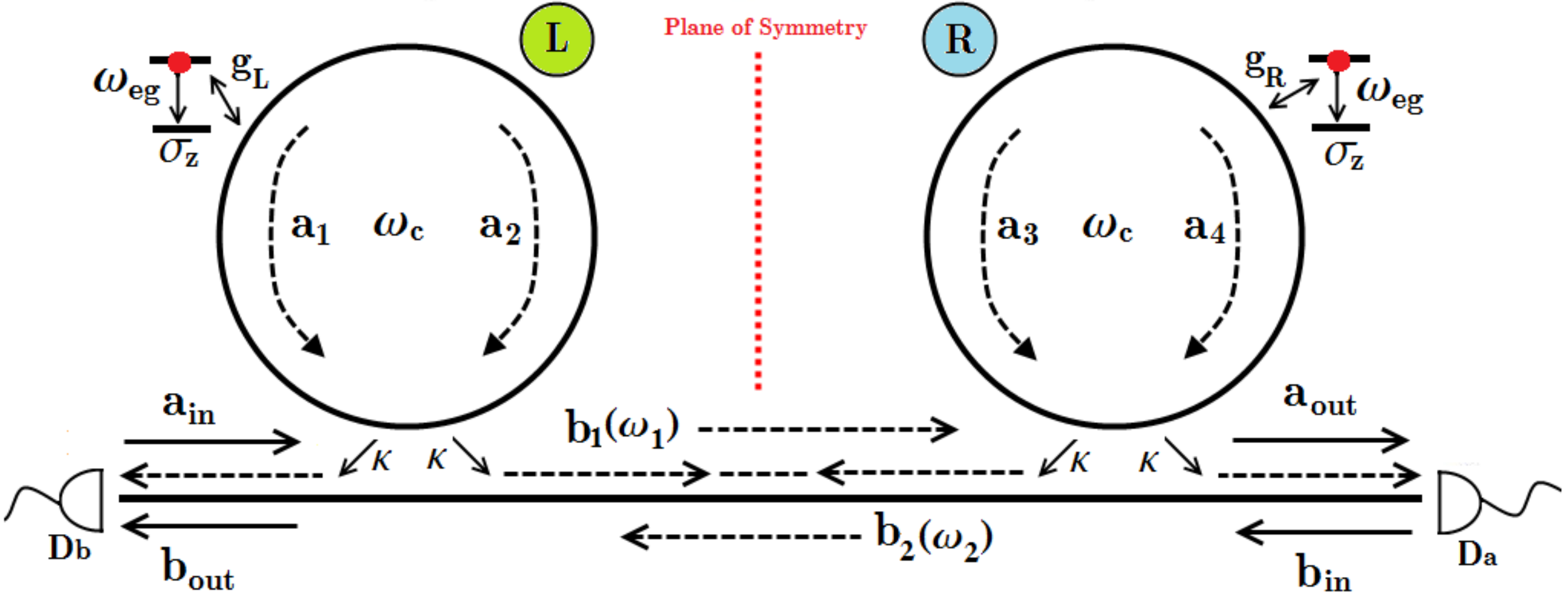}
\captionsetup{
  format=plain,
  margin=1em,
  justification=raggedright,
  singlelinecheck=false
}
  \caption{Two spatially separated atom-cavity systems, and two single-photon detectors. Thanks to the bi-directional coupling between the two cavities, excitations can be transfered between the atom-cavity systems multiple times before being detected. We consider here a mirror-symmetric system, with all coupling constants, decay rates, and resonance frequencies pairwise the same for the left and right atom-cavity systems.
The detectors count photons in the two output modes, described by annihilation operators $\hat{a}_{out}$ and $\hat{b}_{out}$. For further details, see main text.
    }\label{Fig1}
 \end{figure*}
We have two spatially separated atom-cavity systems (referred to as ``left'' or ``L'' and ``right'' or ``R'', respectively) coupled through an optical fiber which is assumed to have two continua of modes (propagating to the left and right, respectively), as shown in FIG.~1. A single photon is generated in each cavity through an initially excited atom (with transition frequency $\omega_{eg}$: both atoms are taken to be identical in the rest of the paper). The spontaneous emission from the atoms is set to zero. In practice one suppresses the effects of spontaneous emission by using three-level atoms in the $\Lambda$ configuration. The excited state can be eliminated adiabatically (it is only off-resonantly coupled), and the resulting description is that of an effective two-level system, where both levels are ground states. 

Due to the atom-cavity coupling (represented by complex coupling coefficients $g_{L}$ and $g_{R}$ for left and right systems, respectively) the emitted photon can excite any one of the two counter propagating cavity modes, which are described by annihilation operators $\hat{a}_{1}$, $\hat{a}_{2}$ for the left cavity and $\hat{a}_{3}$, $\hat{a}_{4}$ for the right cavity. Inside each cavity, both modes are assumed to have the same single resonant frequency $\omega_{c}$.

There are two possibilities for the excitation to leak out of a given cavity. For example, for the left cavity, the photon in the mode $\hat{a}_{2}$ can exit towards the left (at a leakage rate $\kappa$) and will be detected by detector $D_{b}$.
 On the other hand, if the photon is in the mode $\hat{a}_{1}$, then it can escape towards the right (at the same leakage rate $\kappa$), after which it can enter into the right cavity due to the evanescent coupling between fiber and  cavity. It may, alternatively, go straight to the detector $D_a$.
Excitations can shuttle back and forth many times before finally being lost by the system and detected by the two detectors.
 
In our system there is a time delay $\tau$ between the cavities (which is defined in terms of the separation $d$ between cavities as $\tau =d/c$, with $c$ the group velocity of light in the fiber, which is assumed to be constant around the cavities' and atoms' resonant frequencies). Such time delays appear in the context of cascaded quantum networks \cite{carmichael1993quantum,gardiner1993driving} where they are considered arbitrary constants that can be eliminated, since they prove irrelevant to the physics of the problem. But for our system  we cannot so simply ignore the time delay. This is due to the fact that the coupling between system L and R is not unidirectional. From this perspective our model resembles more a quantum feedback network \cite{wiseman2009quantum,gough2009quantum}, with the difference that there is no special part added to the actual system to perform this feedback \cite{petersen2011cascade,vitali2000quantum}. Rather, this happens due to the geometry of the system itself.

Assuming no coupling between the intra cavity modes and applying the standard rotating wave (RWA) and Markov approximations, the Hamiltonian of the global system (atoms, cavities and the fiber) takes the following form:
\begin{widetext}
\begin{equation}\label{H}
\begin{split}
& \hat{H}/\hbar=-\omega_{eg}\hat{\sigma}^{(L)}_{-}\hat{\sigma}^{(L)}_{+}-\omega_{eg}\hat{\sigma}^{(R)}_{-}\hat{\sigma}^{(R)}_{+}+ 
\omega_{c}(\hat{a}_{1}^{\dagger}\hat{a}_{1}+ \hat{a}_{2}^{\dagger}\hat{a}_{2}
+\hat{a}_{3}^{\dagger}\hat{a}_{3}+ \hat{a}_{4}^{\dagger}\hat{a}_{4})
+(g_{L}\hat{a}_{1}^{\dagger}\hat{\sigma}^{(L)}_{-}+g^{\ast}_{L}\hat{a}_{1}\hat{\sigma}^{(L)}_{+}) +(g^{\ast}_{L}\hat{a}_{2}^{\dagger}\hat{\sigma}^{(L)}_{-}\\
&+g_{L}\hat{a}_{2}\hat{\sigma}^{(L)}_{+})+(g_{R}\hat{a}_{3}^{\dagger}\hat{\sigma}^{(R)}_{-}+g^{\ast}_{R}\hat{a}_{3}\hat{\sigma}^{(R)}_{+})+(g^{\ast}_{R}\hat{a}_{4}^{\dagger}\hat{\sigma}^{(R)}_{-} +g_{R}\hat{a}_{4}\hat{\sigma}^{(R)}_{+})
+\int_{-\infty}^{+\infty}\omega_{1} \hat{b}_{1}^\dagger(\omega_{1})\hat{b}_{1}(\omega_{1})d\omega_{1}
+\int_{-\infty}^{+\infty}\omega_{2} \hat{b}_{2}^\dagger(\omega_{2})\hat{b}_{2}(\omega_{2})d\omega_{2}\\
&+i\sqrt{\frac{\kappa}{2\pi}}\int_{-\infty}^{+\infty}\Bigg(\hat{a}_{1}\hat{b}_{1}^{\dagger}(\omega_{1})-\hat{a}_{1}^{\dagger}\hat{b}_{1}(\omega_{1})+\hat{a}_{3}\hat{b}^{\dagger}_{1}(\omega_{1})
-\hat{a}_{3}^{\dagger}\hat{b}_{1}(\omega_{1})\Bigg)d\omega_{1}
+i\sqrt{\frac{\kappa}{2\pi}}\int_{-\infty}^{+\infty}\Bigg(\hat{a}_{2}\hat{b}_{2}^{\dagger}(\omega_{2})-\hat{a}_{2}^{\dagger}\hat{b}_{2}(\omega_{2})
+\hat{a}_{4}\hat{b}^{\dagger}_{2}(\omega_{2})\\
&-\hat{a}_{4}^{\dagger}\hat{b}_{2}(\omega_{2})\Bigg)d\omega_{2}.
\end{split}
\end{equation}
\end{widetext}
Here $\hat{\sigma}^{(L)}_{+},\hat{\sigma}^{(R)}_{+}$ are the atomic raising operators for left and right atoms respectively and $\hat{b}_{1}(\omega_{1})$, $\hat{b}_{2}(\omega_{2})$ are the annihilation operators for two fiber continua. The nonvanishing commutation relations are: $[\hat{\sigma}^{(L)}_{+},\hat{\sigma}^{(L)}_{-}]=\hat{\sigma}^{(L)}_{z}$ and a similar relation for right atom, $[\hat {b}_{i}(\omega_{i}),\hat {b}_{j}^{\dagger}(\omega_{j}))]=\delta(\omega_{i}-\omega_{j})$ $\forall i=1,2$; $j=1,2$ and $[\hat {a}_{i},\hat {a}_{j}^{\dagger}]=\delta_{ij}$ $\forall i=1,2,3,4$; $j=1,2,3,4$. We have chosen the energy of the atomic ground states to be negative (first two terms), such that the initial state has zero energy.

The interaction of the intra cavity modes with the fiber continua makes both left and right systems open and to describe the dynamics of such an open system we now transform to the Heisenberg picture. Following the standard procedure \cite{gardiner2004quantum, carmichael2008statistical} of eliminating continua in the Heisenberg picture and identifying the two input operators corresponding to two continua:
\begin{equation}
\begin{split}
&\hat{a}_{in}(t)=\frac{1}{\sqrt{2\pi}}\int_{-\infty}^{\infty}\hat{b}_{1}(\omega_{1})e^{i\omega_{1}(t-t_{0})}d\omega_{1}\\
&\hat{b}_{in}(t)=\frac{1}{\sqrt{2\pi}}\int_{-\infty}^{\infty}\hat{b}_{2}(\omega_{2})e^{i\omega_{2}(t-t_{0})}d\omega_{2}
\end{split}
\end{equation}
we finally arrive at the following Quantum Langevin's equation for an arbitrary system operator $\hat{X}(t)$ (which can either belong to system L or to system R):
\begin{widetext}
\begin{equation}\label{Lang}
\begin{split}
& \frac{d\hat{X}(t)}{dt}=-\frac{i}{\hbar}[\hat{X}(t),\hat{H}_{s}]
 -[\hat{X}(t),\hat{a}_{1}^\dagger]\Bigg(\frac{\kappa}{2}\hat{a}_{1}+\sqrt{\kappa}\hat{a}_{in}(t)\Bigg)
+\Bigg(\frac{\kappa}{2}\hat{a}_{1}^{\dagger}+\sqrt{\kappa}\hat{a}_{in}^{\dagger}(t)\Bigg)[\hat{X}(t),\hat{a}_{1}]
-[\hat{X}(t),\hat{a}_{3}^\dagger]\Bigg(\frac{\kappa}{2}\hat{a}_{3}\\
&+\sqrt{\kappa}\hat{a}_{in}(t-\tau)\Bigg)+
\Bigg(\frac{\kappa}{2}\hat{a}_{3}^{\dagger}+\sqrt{\kappa}\hat{a}_{in}^{\dagger}(t-\tau)\Bigg)[\hat{X}(t),\hat{a}_{3}]-\kappa[\hat{X}(t),\hat{a}_{3}^{\dagger}]\hat{a}_{1}(t-\tau)+\kappa\hat{a}_{1}^{\dagger}(t-\tau)[\hat{X}(t),\hat{a}_{3}]\\
&-[\hat{X}(t),\hat{a}_{2}^\dagger]\Bigg(\frac{\kappa}{2}\hat{a}_{2}+\sqrt{\kappa}\hat{b}_{in}(t-\tau)\Bigg)+
\Bigg(\frac{\kappa}{2}\hat{a}_{2}^{\dagger}+
\sqrt{\kappa}\hat{b}^{\dagger}_{in}(t-\tau)\Bigg)[\hat{X}(t),\hat{a}_{2}]-[\hat{X}(t),\hat{a}_{4}^\dagger]\Bigg(\frac{\kappa}{2}\hat{a}_{4}+ \sqrt{\kappa}\hat{b}_{in}(t)\Bigg)\\
&+\Bigg(\frac{\kappa}{2}\hat{a}_{4}^{\dagger}+\sqrt{\kappa}\hat{b}_{in}^{\dagger}(t)\Bigg)[\hat{X}(t),\hat{a}_{4}]-\kappa[\hat{X}(t),\hat{a}_{2}^{\dagger}]\hat{a}_{4}(t-\tau)
+\kappa\hat{a}_{4}^{\dagger}(t-\tau)[\hat{X}(t),\hat{a}_{2}].
\end{split}
\end{equation}
\end{widetext}
Here $\hat{H}_{s}$ is the atom-cavity system Hamiltonian, which consists of the discrete terms in Eq.~[\ref{H}]. The above Langevin equation is a generalization of the usual cascaded quantum system Langevin equation \cite{gardiner1993driving,gardiner2004quantum} to include a bidirectional coupling between left and right systems. Corresponding to two input field operators $\hat{a}_{in}$, $\hat{b}_{in}$  appearing in the above equation there are two output operators $\hat{a}_{out}$, $\hat{b}_{out}$ which are related to the input operators and the intra cavity field operators through the input-output relations \cite{carmichael2008statistical,gardiner1985,carmichael1993open} as 
\begin{subequations}
\begin{eqnarray}
\hat{a}_{in}^{(R)}(t)=\hat{a}_{{\rm out}}^{(L)}(t-\tau)=\hat{a}_{{\rm in}}^{(L)}(t-\tau)+\sqrt{\kappa}\hat{a}_{1}(t-\tau),\\
\hat{b}_{in}^{(L)}(t)=\hat{b}_{{\rm out}}^{(R)}(t-\tau)=\hat{b}_{{\rm in}}^{(R)}(t-\tau)+\sqrt{\kappa}\hat{a}_{4}(t-\tau).
\end{eqnarray}
\end{subequations}
Note that the output from one cavity is serving as the input to the other cavity (with the delay time included), so that the coupling is explicitly bidirectional. We have also explicitly included (redundant) L and R superscripts here to make the distinction among the various input and output operators more transparent. The nonvanishing commutation relations among the input operators are given by:
$[\hat{a}_{{\rm in}}(t),\hat{a}_{{\rm in}}^{\dagger}(t')]=\delta(t-t')$, $[\hat{b}_{{\rm in}}(t),\hat{b}_{{\rm in}}^{\dagger}(t')]=\delta(t-t')$. 

If we denote by $\ket{\Psi}$ the initial state of the global system (atoms, cavities and fiber), we have $\hat{a}_{{\rm in}}\ket{\Psi}=0$ and $\hat{b}_{{\rm in}}\ket{\Psi}=0$, as initially there is no photon present. These input operators, therefore, do not contribute to the expectation values of normally ordered observables.

Although the time delay arising from the fiber cannot be ignored due to the feedback mechanism in our system, for the present study we are more interested in the delays caused by the excitations remaining inside the cavities. (In fact, the whole point of using coupled cavity arrays is to store and delay photons inside cavities.) This cavity-induced time delay is on the order of $\kappa^{-1}$ and under the condition that $\kappa\tau<<1 $ we can in fact ignore the trivial delay $\tau$. From now on we are going to focus on this particular regime---the experimentally relevant regime---and we set $\tau\rightarrow 0$ for that reason.

\subsection{Quantum trajectory analysis}
Now we transform back to the Schr\"odinger picture and make use of the Quantum Trajectory Method (or quantum jump method) \cite{carmichael1993open,dum1992monte, molmer1993monte} which is an appropriate formalism for the description of open quantum systems. This analysis applied to the system under study implies that during any (infinitesimally) small time interval we have one of two possibilities: either a photon leaks out of the system and one of the detectors registers it (and so a quantum jump takes place), or the excitation(s) remain inside the system and no jump is recorded. The next subsections are devoted to the detailed study of both these situations.

\subsubsection{Occurrence of a jump}
In the Quantum Trajectory Method, photodetection at the output ports is described by the output operators (also called jump operators in this context), which in our case are denoted by $\hat{J}_{a}=\hat{a}_{{\rm out}}$ and $\hat{J}_{b}=\hat{b}_{{\rm out}}$. Detector $D_{a}$ detects the field $\hat{a}_{{\rm out}}$ and $D_{b}$ detects the field $\hat{b}_{{\rm out}}$ (see Fig.[\ref{Fig1}]). The detection events happen at random times with certain probabilities determined by the jump/output operators $\hat{J}_j$ for $j=a,b$, and by the current state $\ket{\psi}$. During an infinitesimal time interval $[t,t+dt]$ the detection probability is given by
\begin{equation}\label{pi}
P_{j}(t)={\bra{\psi}}\hat{J_{j}}^{\dagger}
\hat{J_{j}}{\ket{\psi}}dt=:\Pi_j dt,
\end{equation}
for $j=a,b$. After one jump is recorded we have to reset the state of the system according to the transformation:
\begin{equation}
\ket{\psi}\mapsto
\frac{\hat{J}_j\ket{\psi}}{\sqrt{\Pi_j}}.
\end{equation}
The normalization factor $\Pi_j$ appearing here is in fact the probability density defined in Eq.~(\ref{pi}).

\subsubsection{Non-unitary evolution}
According to the Quantum Trajectory Method, when no detector clicks, the system dynamics follows a non-unitary evolution described by a non-unitary Schr\"odinger equation:
\begin{equation}\label{NUSE}
i\hbar\frac{d\ket{\tilde{\psi}(t)}}{dt}=\hat{H}_{NH}{\ket{\tilde{\psi}(t)}}.
\end{equation}
The ``Non-Hermitian Hamiltonian'' $\hat{H}_{NH}$ appearing in the above equation turns out to be the sum of the standard (Hermitian) system Hamiltonian (Eq.~[\ref{H}]) and an anti-Hermitian term constructed from the jump operators, such that
\begin{equation}\label{NHH}
\hat{H}_{NH}=\hat{H}_{s}-i\sum_{j=a,b}\hat{J}_j^\dagger\hat{J}_j/2.
\end{equation}
The unnormalized ket $\ket{\tilde{\psi}(t)}$ is called the ``No-Jump state," which is a pure state whose norm decays in time. (It is normalized back to unity upon the occurrence of a quantum jump.)
It can be written as a linear combination of all the different possibilities of finding all excitations in the system that have not been detected yet.

\section{Two-photon quantum interference effects}
In this Section we will analyze two-photon interference effects. In particular, we study whether the probability to detect the two photons in the same detector differs from the probability to detect them in different detectors.
We consider two cases: first a case of mere theoretical significance where we compare joint detection probabilities in a small time interval (so the photons are detected at the same time), and second a case of experimental relevance where one records at what detectors and at what times the two photons were detected.

Some aspects of  single-photon transmission could be derived using a semi-classical approach (see for instance \cite{srinivasan2007mode}, \cite{elyutin2012interaction}).
Here, on the other hand, we are interested in interference of the Hong-Ou-Mandel type, which cannot be explained semi-classically \cite{mandel}, and we thus
follow the procedure outlined in the preceding Section. The ``No-Jump state'' $\ket{\tilde{\psi}(t)}$ describing the situation where neither excitation has been detected yet,  consists of a superposition of 19 different states, corresponding to the 19 different ways of finding the two excitations in the different parts of the system (2 atoms and 4 cavity modes; the fiber continuum modes have been eliminated from the picture). We write 
\begin{equation}\label{nojumptotal}
\begin{split}
&\ket{\tilde{\psi}(t)}=c_{1}(t)\ket{e_{1}00,e_{2}00}+c_{2}(t)\ket{e_{1}10,g_{2}00}\\
&+c_{3}(t)\ket{e_{1}01,g_{2}00}+c_{4}(t)\ket{e_{1}00,g_{2}10}+c_{5}(t)\ket{e_{1}00,g_{2}01}\\
&+c_{6}(t)\ket{g_{1}10,e_{2}00}+c_{7}(t)\ket{g_{1}01,e_{2}00}+c_{8}(t)\ket{g_{1}00,e_{2}10}\\
&+c_{9}(t)\ket{g_{1}00,e_{2}01}+c_{10}(t)\ket{g_{1}20,g_{2}00}+c_{11}(t)\ket{g_{1}02,g_{2}00}\\
&+c_{12}(t)\ket{g_{1}00,g_{2}20}+c_{13}(t)\ket{g_{1}00,g_{2}02}+c_{14}(t)\ket{g_{1}11,g_{2}00}\\
&+c_{15}(t)\ket{g_{1}10,g_{2}10}+c_{16}(t)\ket{g_{1}10,g_{2}01}+c_{17}(t)\ket{g_{1}01,g_{2}10}\\
&+c_{18}(t)\ket{g_{1}01,g_{2}01}+c_{19}(t)\ket{g_{1}00,g_{2}11}.
\end{split}
\end{equation}
The notation we used here is as follows: the first slot in the ket is the state of the left atom and the next two slots display the number of photons in the modes of the left cavity. The remaining three slots are for the right system with the atomic and cavity states ordered in the same way. 
\begin{figure*}
\begin{center}
\begin{tabular}{cccc}
\subfloat{\includegraphics[width=5cm,height=4cm]{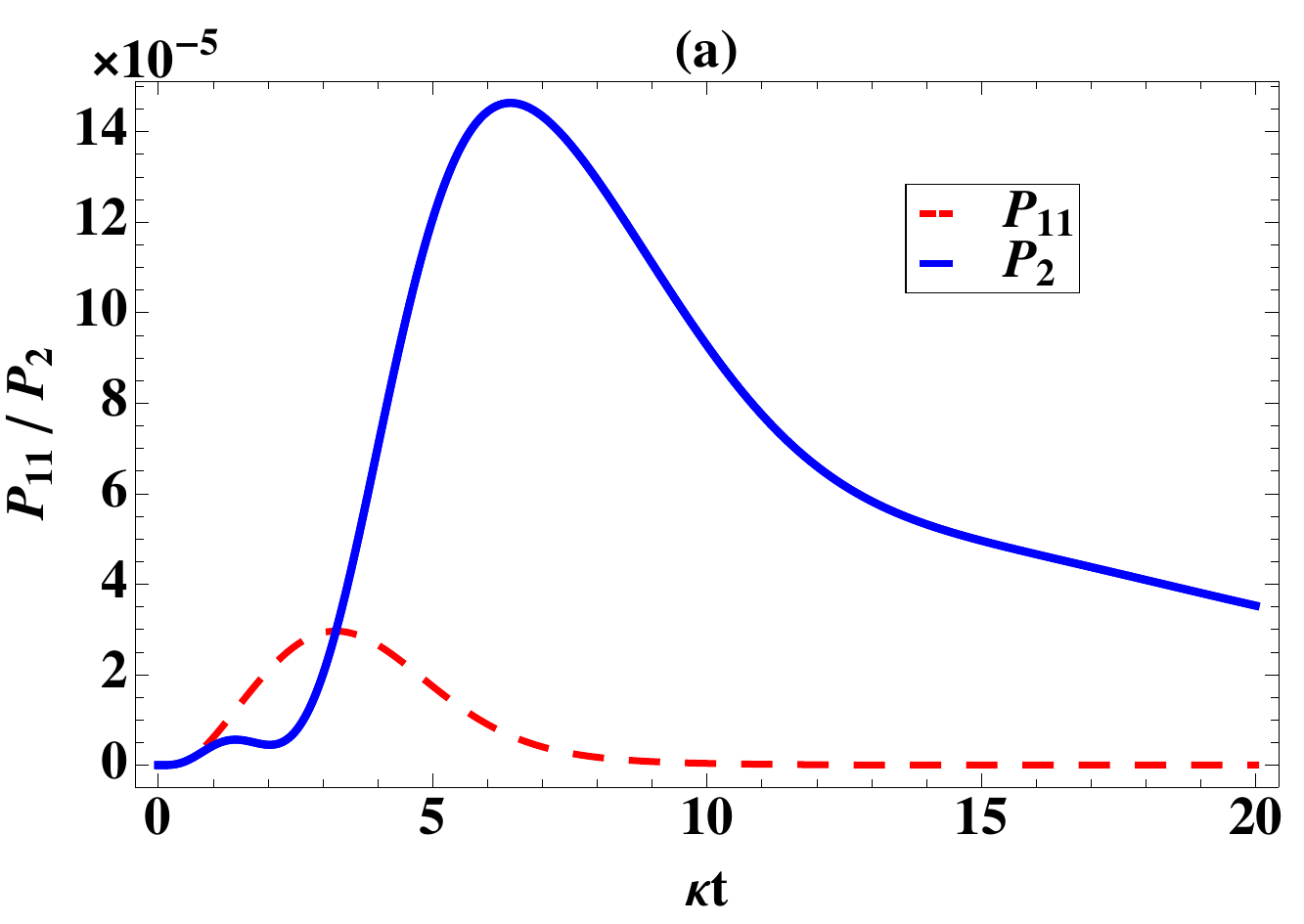}} & 
\subfloat{\includegraphics[width=5cm,height=4cm]{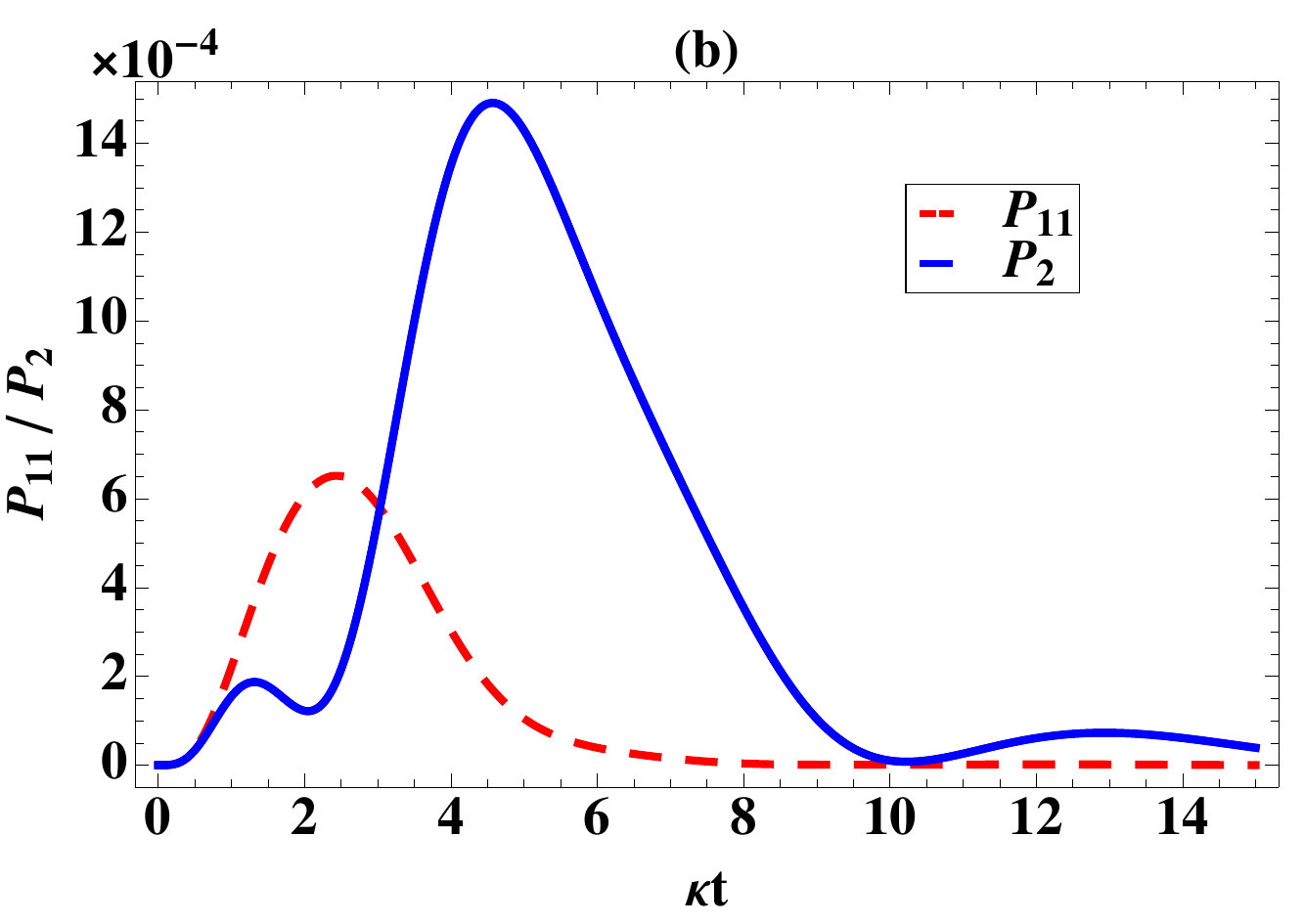}}& 
\subfloat{\includegraphics[width=5cm,height=4cm]{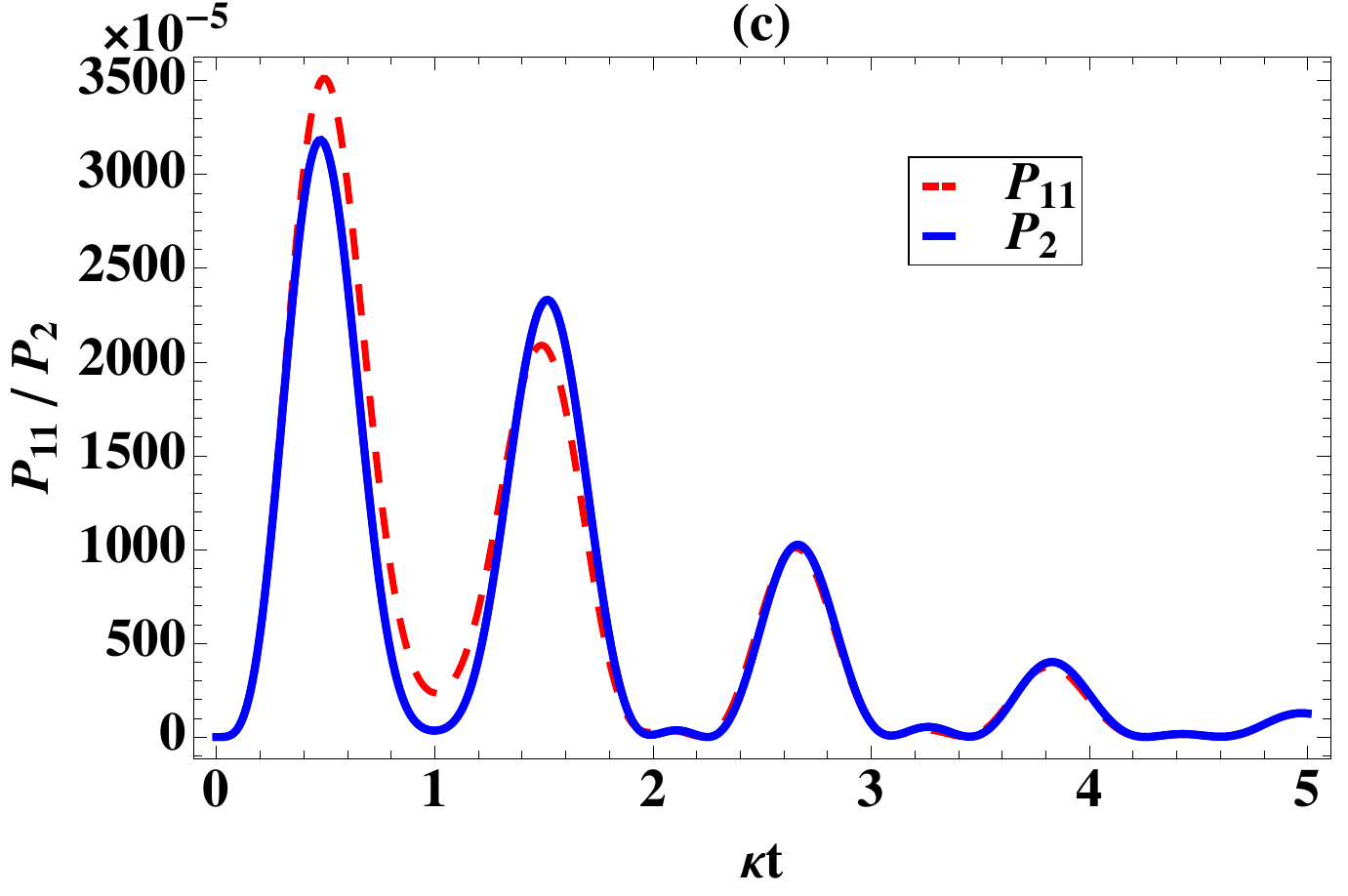}}\\
\end{tabular}
\captionsetup{
  format=plain,
  margin=1em,
  justification=raggedright,
  singlelinecheck=false
}
\caption{Joint probability densities of detecting photons at the output ports as functions of time, for three values of $|g|/\kappa$.  In
(a) we choose $|g|/\kappa = 0.1, \Delta/\kappa=0.5, {\rm \delta T = 0.1\kappa^{-1} }$, and in 
(b) we choose  $|g|/\kappa=0.25$ with all other parameters the same as in part (a). In both graphs it is considerably more likely to detect the two photons at the same output, rather than at different outputs, reminiscent of the Hong-Ou-Mandel effect. One does notice this effect slightly diminishing when going from (a) to (b). In (c) we choose $|g|/\kappa=2$ and all other parameters the same as in part (a). Entering this regime we notice that the probability densities display Rabi oscillations. The main point of (c) is to show that in this nonlinear regime, the probabilities to detect photons at one and the same detector or at two different detectors have become almost equal. The HOM interference effect disappears.}\label{Fig2}
\end{center}
\end{figure*}
\subsection{Photons detected at the same time}
We study the interference effects in our system by first calculating the joint probabilities of detecting the two photons at the output ports. Here we remind the reader that we are working in the regime where trivial fiber delays are neglected, and so one type of interference (of a theoretical nature) can be studied by considering the equal-time probability densities. We thus compare 
\begin{equation}
\begin{split}\label{P2}
&\text{Probability density of getting two clicks at the same }\\
& \text{time t at detector $D_{a}$}\equiv P_{2}= \langle \tilde{\psi}(t) \vert \hat{a}_{out}^{\dagger 2}\hat{a}_{out}^{2} \vert \tilde{\psi}(t) \rangle {\rm \delta T}\\
&\hspace{20mm} =\kappa^{2}\Bigg\vert \sqrt{2}c_{10}(t)+\sqrt{2}c_{12}(t)+2c_{15}(t)\Bigg\vert^{2}{\rm \delta T}
\end{split}
\end{equation}
(with $\delta T$ is a very small time interval compared to the cavity leakage time $\kappa^{-1}$) 
with
\begin{equation}\label{P11}
\begin{split}
&\text{Probability density of getting one click at detector $D_{a}$}\\
& \text{and the other at detector $D_{b}$ at the same time t}\\
& \hspace{5mm}\equiv P_{11}= \langle \tilde{\psi}(t) \vert \hat{b}_{out}^{\dagger}\hat{a}_{out}^{\dagger}\hat{a}_{out}\hat{b}_{out} \vert \tilde{\psi}(t) \rangle {\rm \delta T}\\
& \hspace{15mm}=\kappa^{2}\Bigg\vert c_{14}(t)+c_{16}(t)+c_{17}(t)+c_{19}(t)\Bigg\vert^{2}{\rm \delta T}
\end{split}
\end{equation}
We will call the latter the ab/ba detection density and the former the aa/bb detection density (here we use the fact that because of the mirror symmetry we imposed, the joint probability of getting two clicks at detector a is the same as that of getting two clicks at detector b). 

We plot both densities in FIG.~2 as functions of time, for three different values of $|g|/\kappa$. The HOM interference effect is plainly visible in FIG.~2(a), where almost all photons detected at equal times, were detected at the same detector. If we increase the nonlinear interaction between atoms and photons, we do see this effect disappear. In the strong coupling regime (FIG.~2(c)), we see two almost overlapping curves. Photons detected at the same time arrive at random detectors, and there is no bunching. 
\begin{figure*}
\begin{center}
\begin{tabular}{cccc}
\subfloat{\includegraphics[width=4.1cm,height=6cm]{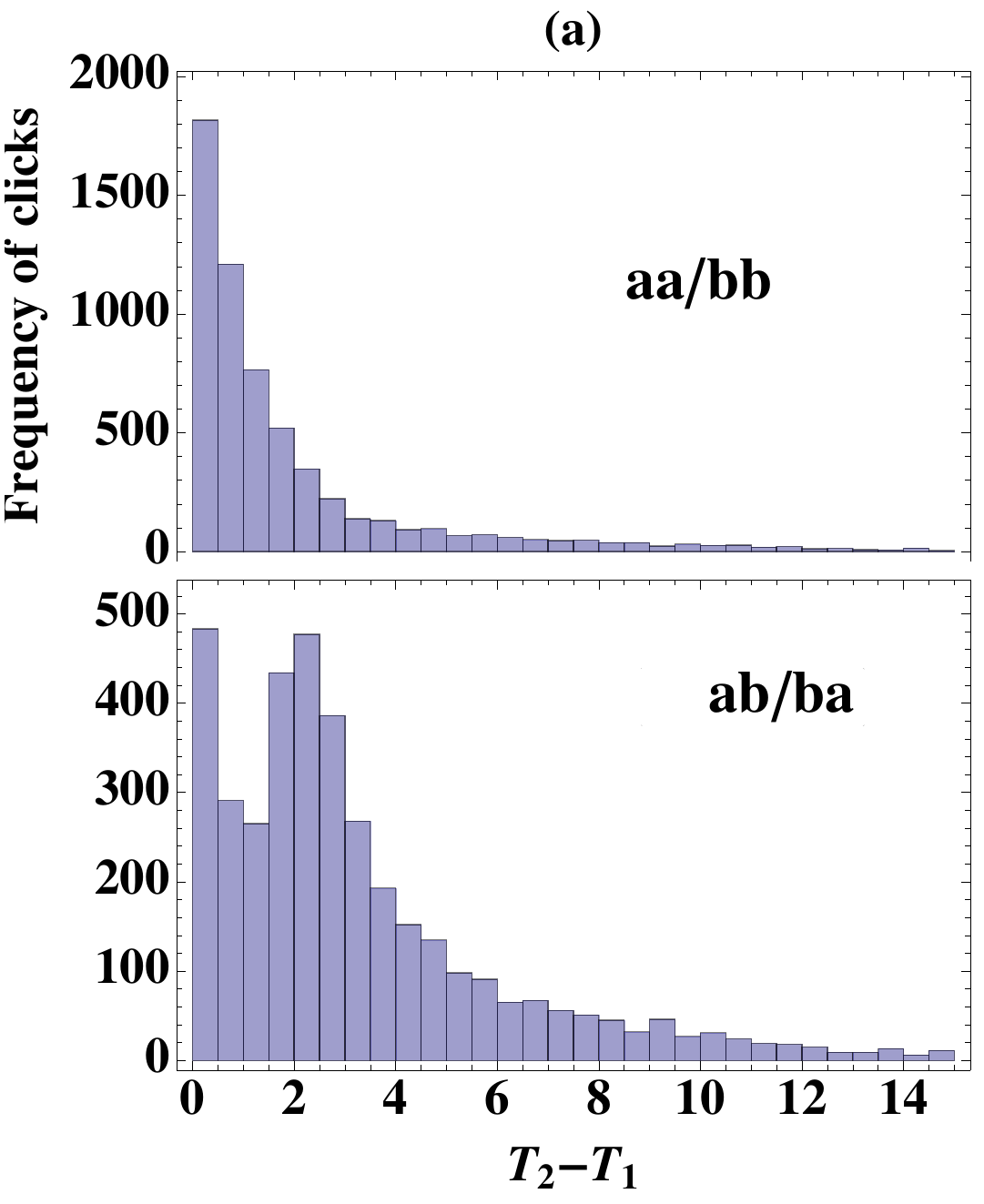}} & 
\subfloat{\includegraphics[width=4.1cm,height=6cm]{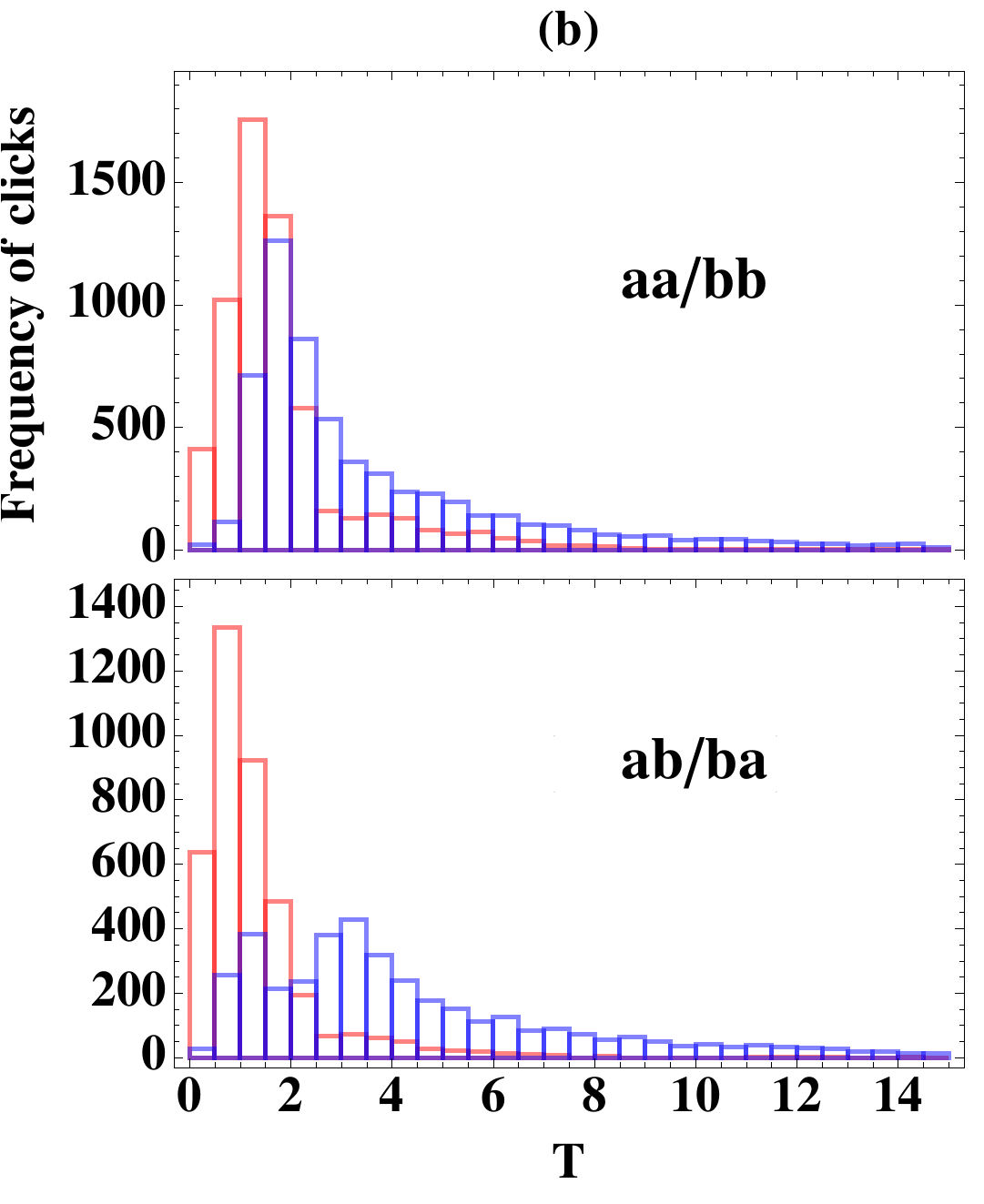}}& 
\subfloat{\includegraphics[width=4.1cm,height=6cm]{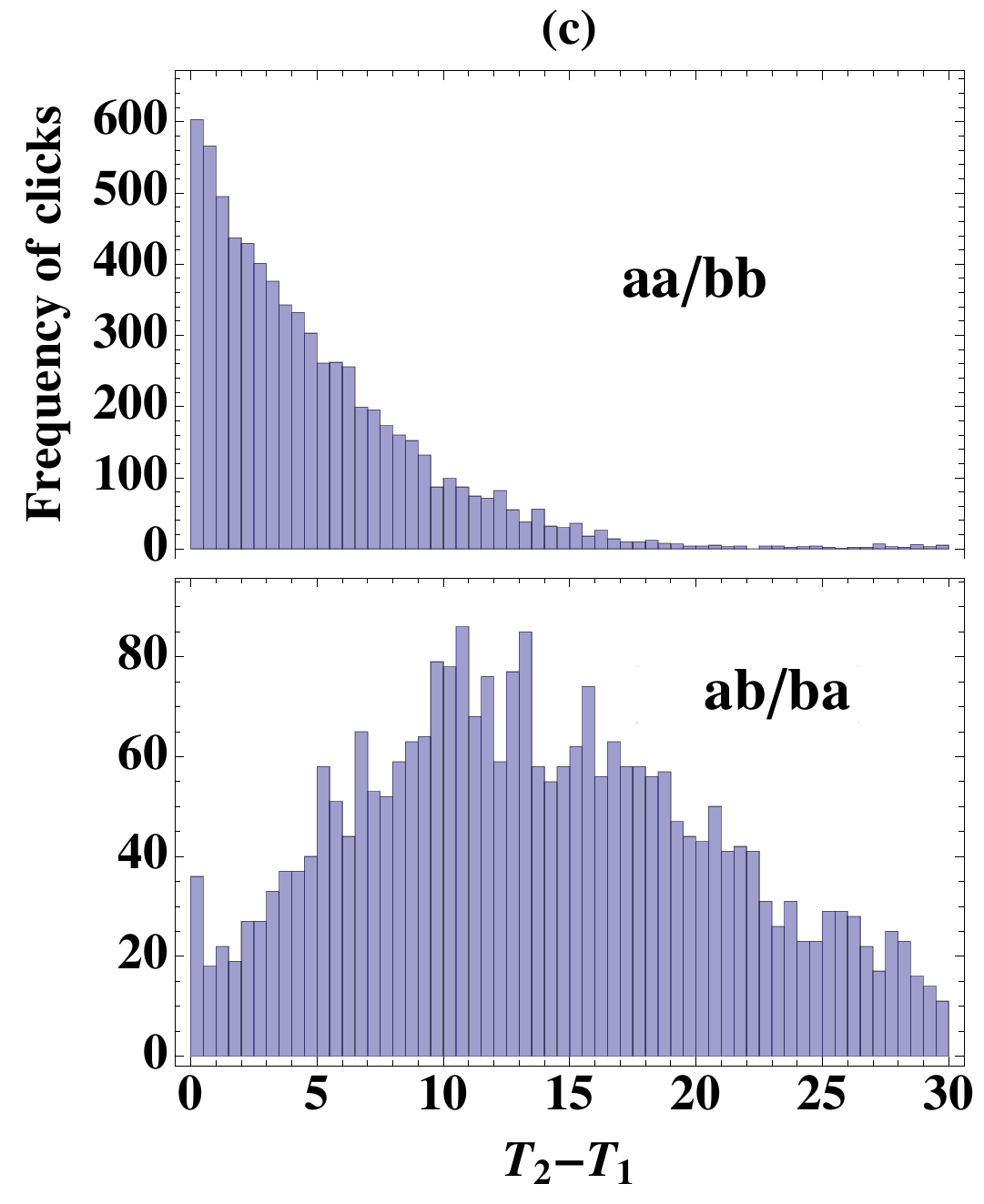}}&\\ 
\subfloat{\includegraphics[width=4.1cm,height=6cm]{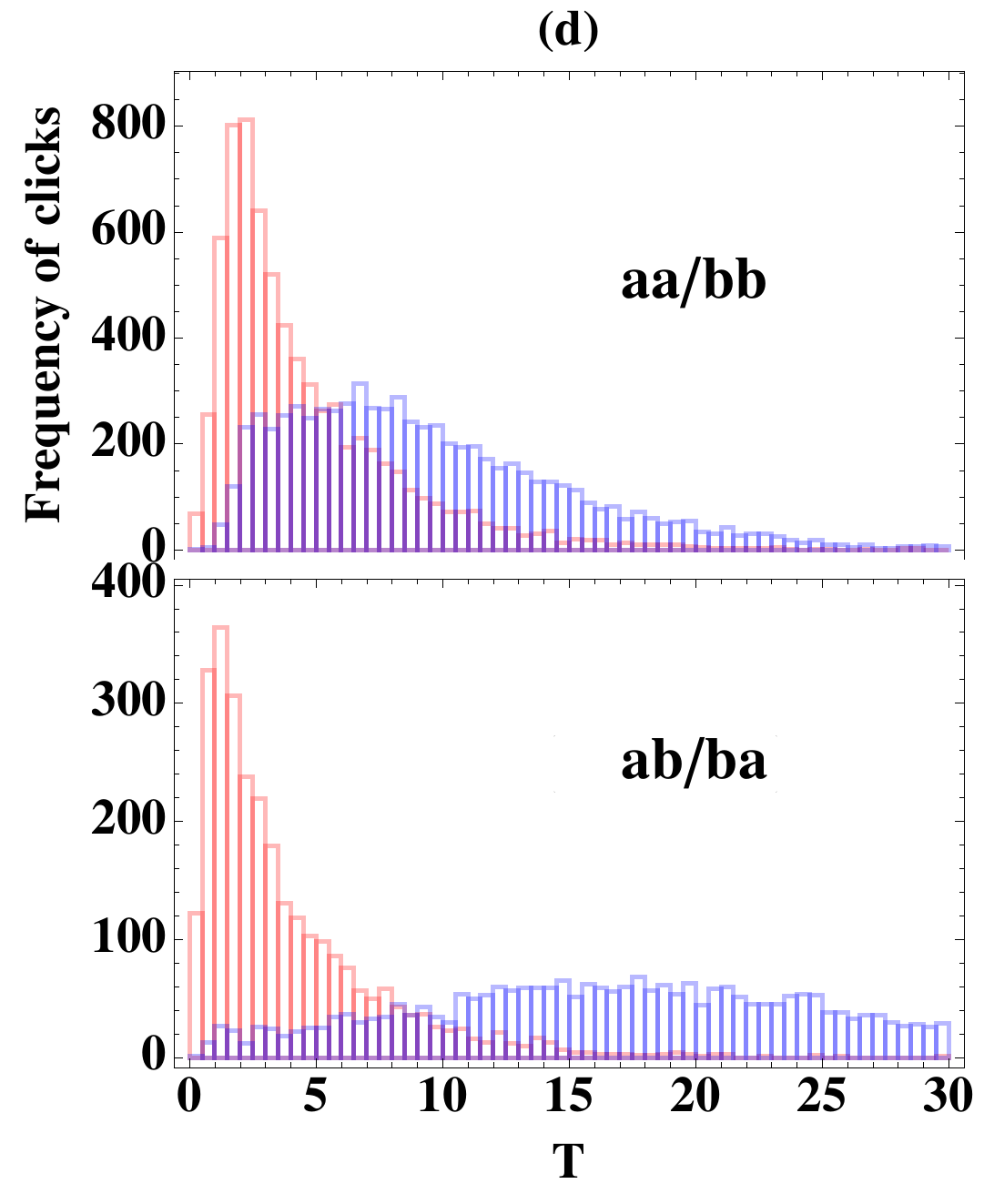}}&
\subfloat{\includegraphics[width=4.1cm,height=6cm]{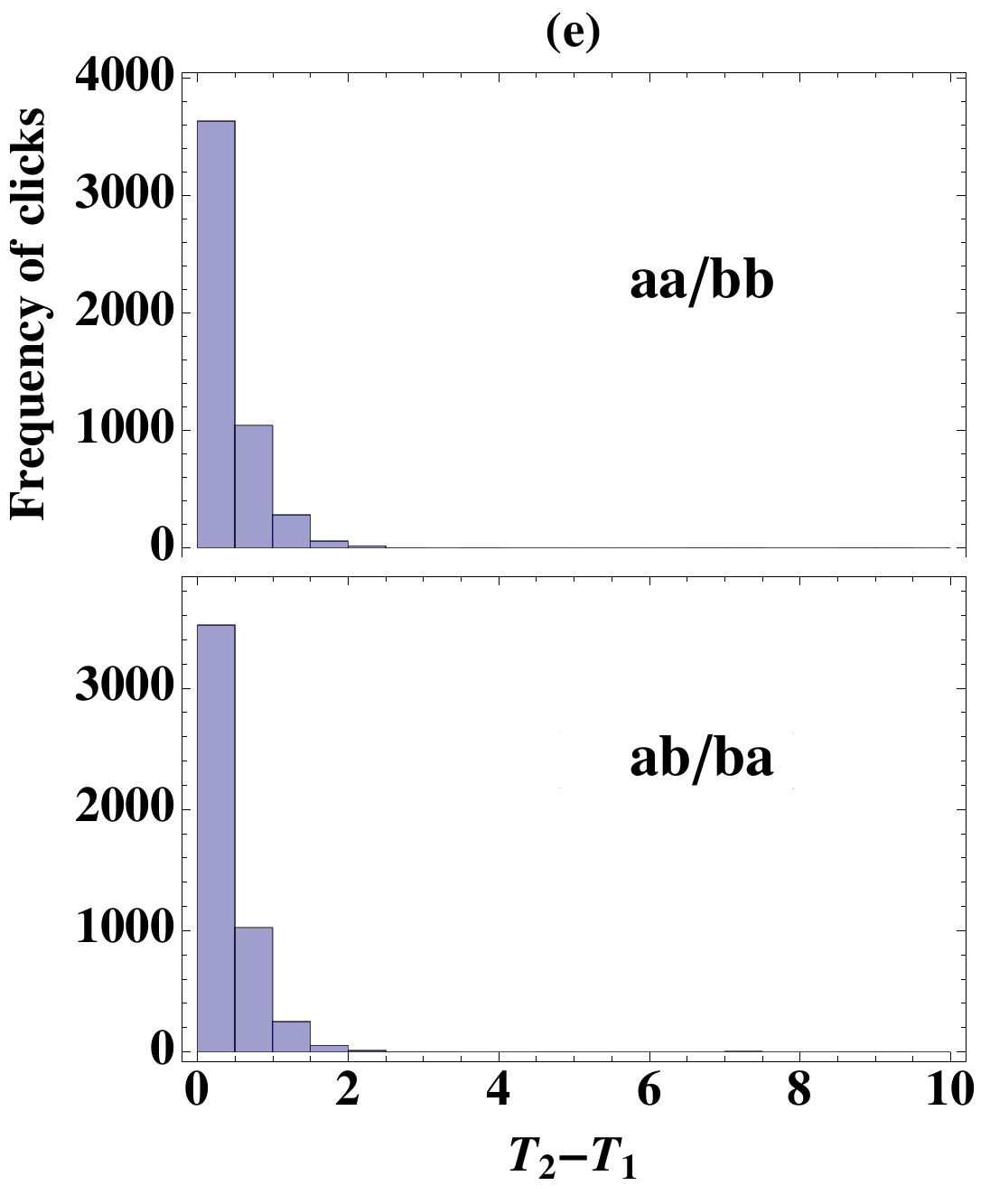}}& 
\subfloat{\includegraphics[width=4.1cm,height=6cm]{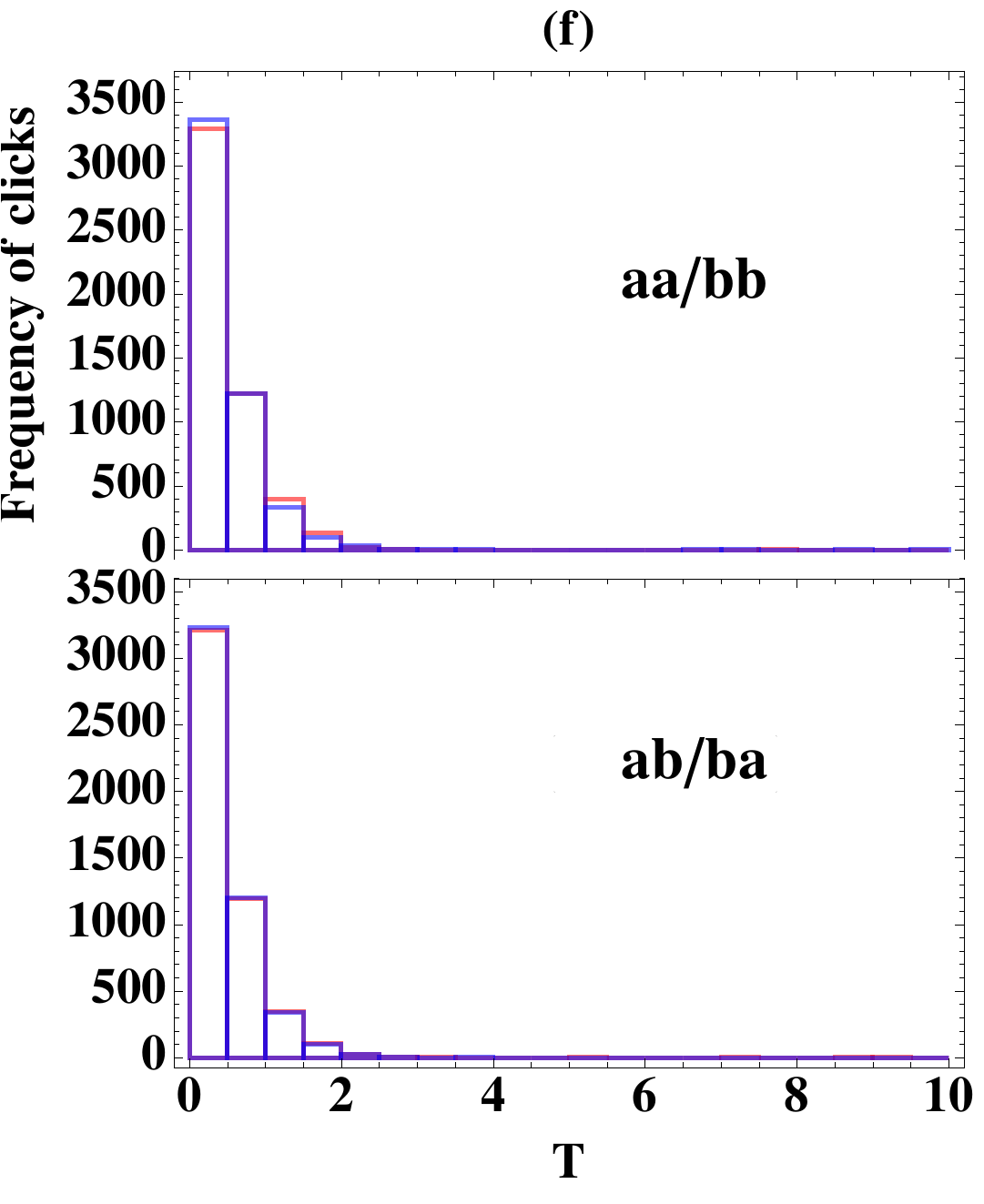}}
\end{tabular}
\captionsetup{
  format=plain,
  margin=1em,
  justification=raggedright,
  singlelinecheck=false
}
\caption{Frequency histograms of detection time differences (left column) and detection time (right column) in the weak coupling regime: For FIGs.~3(a),(b) we chose  $g/\kappa=0.25$, and for FIGs. 3(c),(d) $g/\kappa=0.1$.
We enter the strong coupling regime in FIGs.~ 3(e),(f) where $g/\kappa=5$. All other parameters are the same as used in FIG.~2. The widths of the bins of the histograms are chosen to be $0.5\kappa$. The left figure in each column refers to the time difference $(T_{2}-T_{1})$ and the right figure in all columns displays $T_{1}$ and $T_{2}$ histograms plotted on same scale. In all of the numerical Monte Carlo simulations we have chosen a time step of $0.1\kappa^{-1}$.}\label{Fig3}
\end{center}
\end{figure*}
\subsection{Photons detected at arbitrary times}
The probability of getting two clicks at more or less the same time (as defined in Eq.~[\ref{P2}] and Eq.~[\ref{P11}] respectively) will be very small. A more likely event is that we detect the photons at different times.
We address this situation by considering a simulation of a feasible experiment. The experiment records at what times which detectors click, and the analysis of the data is then supposed to reveal the presence of quantum interference. The latter ought to be manifested in differences between the distributions of waiting times between clicks at the same detector and waiting times between clicks at different detectors.

We performed a Quantum Monte Carlo simulation consisting of over 10,000 trajectories (with a numerical evolution time step of $0.1\kappa$) and we recorded the times at which the two photons are detected at the outputs. We use the following convention: time $T_{1}$ indicates the time of arrival of the first detected click, and $T_{2}$ that of the second.  By this definition $T_{2}>T_{1}$. 

In FIG. 3 we have plotted the frequency histograms of the individual detection, as well as the time differences $T_{2}-T_{1}$, for three different set of parameters. In the upper four plots (a)--(d) we used weak coupling regime parameters with $g/\kappa = 0.25$ for FIG. 3(a),-(b) and $g/\kappa = 0.1$ for FIG. 3(c),-(d). In the bottom plots (FIG. 3(e),-(f)) we have entered the strong coupling regime by taking $g/\kappa = 5$. It turns out from the simulations corresponding to FIGs. 3(a)-(b), 62 percent of the trajectories lead to clicks at the same detector, while in 38\% of the cases photons are detected at different detectors. We notice that working in the even weaker regime ($g/\kappa = 0.1$, FIGs.3(c)-(d)) results in further increase in the same detector detection to 71\%. This imbalance of detection events is a clear indication of HOM type of interference in the weak coupling regime. 

Here we wish to note that for our coupled cavity system (as shown in FIG. 1) the maximum limit on perfect HOM interference is achieved when detection at the same detector happens with 75\% probability.
To understand this maximum limit, we notice that in our coupled cavity system HOM interference can occur only for the situations when both photons will be leaked towards the central region in between the two cavities (through modes corresponding to annihilation operators $\hat{a}_{1}$ and $\hat{a}_{4}$). This happens with 75\% probability which sets this upper limit.

From the weak coupling regime plots in FIG. 3, it is clear that for the $aa/bb(T_{1},T_{2})$ detections the time difference is typically shorter than that for the $ab/ba(T_{1},T_{2})$ detections (there is no bunching effect at all in the latter, but there is in the former). We also note the presence of destructive interference in the $ab/ba(T_{1},T_{2})$ detection case around a time $2\kappa^{-1}$ in FIG. 3(a). 
 
For the strong coupling regime histograms (FIG. 3(e),-(f)) we find that the probability of aa/bb($T_{1},T_{2}$) detection events reduces to 51\%  with a corresponding increase in ab/ba($T_{1},T_{2}$) events to 49\% of trajectories. Moreover, photon bunching becomes less pronounced in aa/bb($T_{1},T_{2}$) histograms compared to ab/ba($T_{1},T_{2}$) plots. This behavior shows that in the strong coupling regime, the nonlinear interactions between the two photons and the two atoms become the dominant factor, and the nonlinearities destroy the HOM interference. 

\section{Conclusions}
We studied in some detail interference effects between two photons in a coupled cavity system. The interference is of the Hong-Ou-Mandel type, and our calculations showed that the two photons are in general more likely to be detected by one and the same detector, rather than by two different detectors.
The destructive interference between different pathways leading to the photons ending up in different detectors thus survives, at least partially, both nonlinear optics effects (due to the presence of atoms in our cavities) and spectral filtering by the resonant cavities. But the larger the nonlinearity, the more the destructive interference disappears, until it is destroyed quite completely in the strong-coupling regime when photons arrive at the two detectors randomly.

\bibliography{article}
\end{document}